\documentclass[a4paper,fleqn,usenatbib]{mnras}

\usepackage{newtxtext,newtxmath}

\usepackage[T1]{fontenc}
\usepackage{ae,aecompl}

\usepackage{graphicx}	
\usepackage{amsmath}	
\usepackage{amssymb}	

\title[The host galaxies of FeLoBAL quasars]{The host galaxies of FeLoBAL quasars at z$\sim$0.9 are not dominated by recent major mergers}

\author[Villforth et al.]{
C. Villforth,$^{1, 2}$\thanks{E-mail: c.villforth@bath.ac.uk}
H. Herbst,$^{3}$
F. Hamann,$^{4}$
T. Hamilton,$^{5}$
C. Bertemes,$^{1}$
A. Efthymiadou,$^{1}$\newauthor
and T. Hewlett$^{2}$
\\
$^{1}$University of Bath, Department of Physics, Claverton Down, Bath, BA2 7AY, UK\\
$^{2}$SUPA, School of Physics and Astronomy, University of St Andrews, North Haugh, KY16 9SS St Andrews, UK\\
$^{3}$Department of Astronomy, University of Florida, Gainesville, FL 32611, USA\\
$^{4}$Department of Physics \& Astronomy, University of California, Riverside, CA 92507, USA\\
$^{5}$Department of Physics, Shawnee State University, 940 Second Street, Portsmouth, OH 45662, USA
}

\date{Accepted XXX. Received YYY; in original form ZZZ}

\pubyear{2018}

\begin{document}
\label{firstpage}
\pagerange{\pageref{firstpage}--\pageref{lastpage}}
\maketitle

\begin{abstract}
Theoretical models have suggested an evolutionary model for quasars, in which most of luminous quasars are triggered by major mergers. It is also postulated that reddening as well as powerful outflows indicate an early phase of activity, close to the merger event. We test this model on a sample of quasars with powerful low ionization outflows seen in broad Iron absorption lines (FeLoBAL). This sample of objects show strong reddening in the optical and fast ($\sim$0.1c) high column density outflows. We present HST WFC3/IR F160W imaging of 10 FeLoBAL host galaxies at redshifts z$\sim$0.9 ($\lambda_{rest}\sim8500\AA$). We compare the host galaxy morphologies and merger signatures of FeLoBALs to luminous blue non-BAL quasars from \citet{villforth_host_2017} of comparable luminosity, which show no excess of merger features compared to inactive control samples. If FeLoBAL quasars are indeed in a young evolutionary state, close in time to the initial merging event, they should have strong merger features. We find that the host galaxies of FeLoBAL quasars are of comparable luminosity to the host galaxies of optical quasars and show no enhanced merger rates. When looking only at quasars without strong PSF residuals, an enhancement in disturbed and merger rates is seen. While FeLoBAL hosts show weak enhancements over a control of blue quasars, their host galaxies are not dominated by recent major mergers.
\end{abstract}




\section{Introduction}

It has been shown that all massive galaxies host central supermassive black holes (SMBHs) \citep[][and references therein]{kormendy_coevolution_2013}. The empirical correlation between the SMBH mass and velocity dispersion of the bulge component of the host (M-sigma relationship) \citep[e.g.][]{gebhardt_relationship_2000,gultekin_m-_2009} as well as other galaxy properties, such as absolute luminosity and mass \citep[e.g.][and references therein]{novak_correlations_2006,kormendy_coevolution_2013} suggest coevolution of SMBHs and their host galaxies. This is also supported by the similar redshift peaks in quasar\footnote{We use both the term quasar and Active Galactic Nucleus (AGN). In this paper, quasar refers to luminous sources ($log(L_{bol}) \geqslant 45$), while AGN refers to active black holes irrespective of luminosity.} activity and star formation rates \citep[e.g.][]{shaver_decrease_1996,madau_star_1998}. Based on these observations, some models have suggested that merger of gas rich galaxies that triggers both a starburst and accretion on to the central SMBH \citep{sanders_ultraluminous_1988,di_matteo_energy_2005} drive the observed colevolution. In these models, the quasar is believed to be initially obscured by the large amounts of gas/dust and appears red. Optically blue quasars are thought to appear once quasar driven outflows have cleared the surrounding gas and dust \citep[e.g.][]{hopkins_cosmological_2008}.

While this evolutionary picture is popular, observational evidence remains mixed. While galaxy interactions are found to increase the incidence of black hole activity \citep[e.g.][]{ellison_galaxy_2013,ellison_galaxy_2015,koss_merging_2010,satyapal_galaxy_2014}, a wide range of studies analyzing the incidence of mergers in AGN have found that their hosts do not show an increased merger rate \citep{cisternas_bulk_2011,kocevski_candels:_2012,villforth_morphologies_2014,villforth_host_2017,hewlett_redshift_2017,mechtley_most_2016}.
Some samples of AGN, such as radio-selected AGN \citep[e.g.][]{ramos_almeida_are_2011,chiaberge_radio_2015}, heaviliy reddened AGN \citep{urrutia_evidence_2008,glikman_major_2015} or X-ray obscured AGN \citep{kocevski_are_2015} however have shown merger rates in excess of either unobscured AGN or control samples. This raises the question of which AGN samples might be associated with an early phase of activity and therefore more closely associated with the triggering event. 

We analyze the host galaxies of a sample of FeLoBAL quasars and compare them to a sample of blue quasars from \citet{villforth_host_2017}. FeLoBALs, which have Fe II and other low-ionization broad absorption lines (LoBALs) \citep{becker_first_1997,hall_unusual_2002}, have been proposed as transition objects based on their typically dust-reddened colors \citep{dunn_determining_2015} and large far-IR luminosities (indicating high star formation rates (SFRs)) \citep{farrah_extraordinary_2010,farrah_direct_2012}, although \citet{violino_scuba-2_2016} found no enhanced star formation rates when compared to the general quasar population. Samples of extremely red quasars, which show high incidences of FeLoBALs, have also been found to have extremely high merger rates \citep{glikman_first-2mass_2012,urrutia_evidence_2008}. FeLoBALs also drive outflows that may be powerful enough to disrupt star formation and drive a galaxy-wide blowout \citep[e.g.][]{moe_quasar_2009,dunn_quasar_2010,faucher-giguere_physics_2012}. However, there is limited understanding of how the host galaxies of FeLoBALs compare to those of normal AGN \citep[see][for a study of 4 FeLoBAL hosts]{lawther_hubble_2018}.

We present Wide Field Camera 3 IR (WFC3/IR) F160W (broad H-band) imaging of 10 FeLoBALs at redshifts z$\sim$0.9 ($\lambda_{rest} \sim$ 8500\AA, tracing stellar mass). We compare the host galaxy morphologies and merger signatures of FeLoBAL quasars with normal luminous blue quasars to test the evolutionary model of FeLoBALs. 
We introduce the sample in Section \ref{S:sample}, observations and data reductions are described in Section \ref{S:data}. Due to emission from the AGN itself, image decomposition is required to recover the host galaxies, as outlined in Section \ref{S:psf}. The morphological analysis is presented in Section \ref{S:visclass}. We discuss the results in Section \ref{S:discussion}, followed by Conclusions in Section \ref{S:end}. All magnitudes are in AB. The cosmology used is $H_{0}=70 \textrm{km}\ \textrm{s}^{-1}, \Omega_m=0.3, \Omega_{\Lambda}=0.7$.

\begin{figure}
\includegraphics[width=\columnwidth]{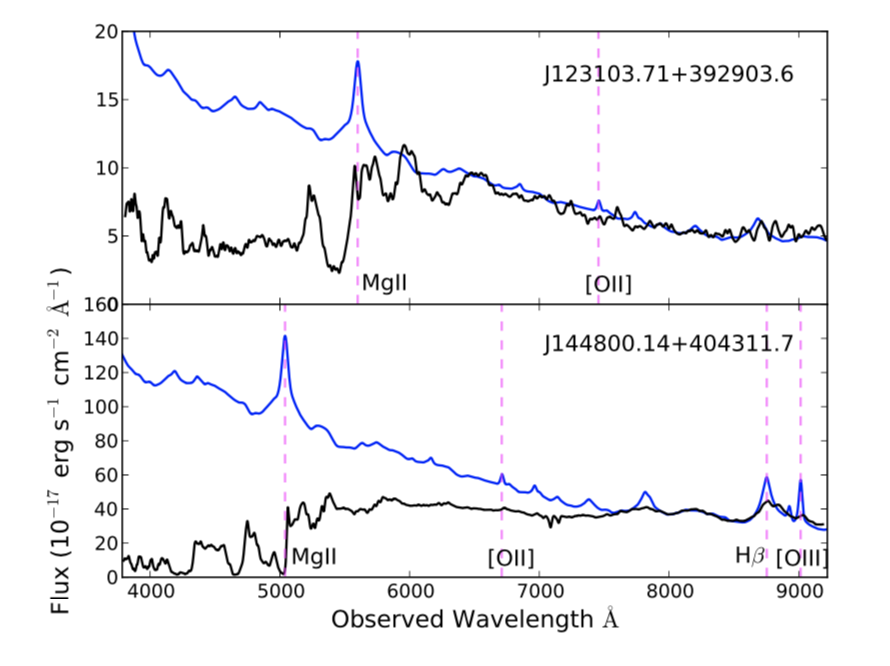}
\caption{Observed-frame SDSS spectra of two representative FeLoBAL quasars in our sample (black curves) compared to a standard non-BAL quasar composite \citep[blue curves;][]{vanden_berk_composite_2001}. The locations of some prominent emission lines are marked by dashed vertical lines. The Mg II and Fe II BALs are seen blueward of the Mg II emission line.}
\label{fig:felobalspec}
\end{figure}

\begin{figure*}
\includegraphics[width=\textwidth]{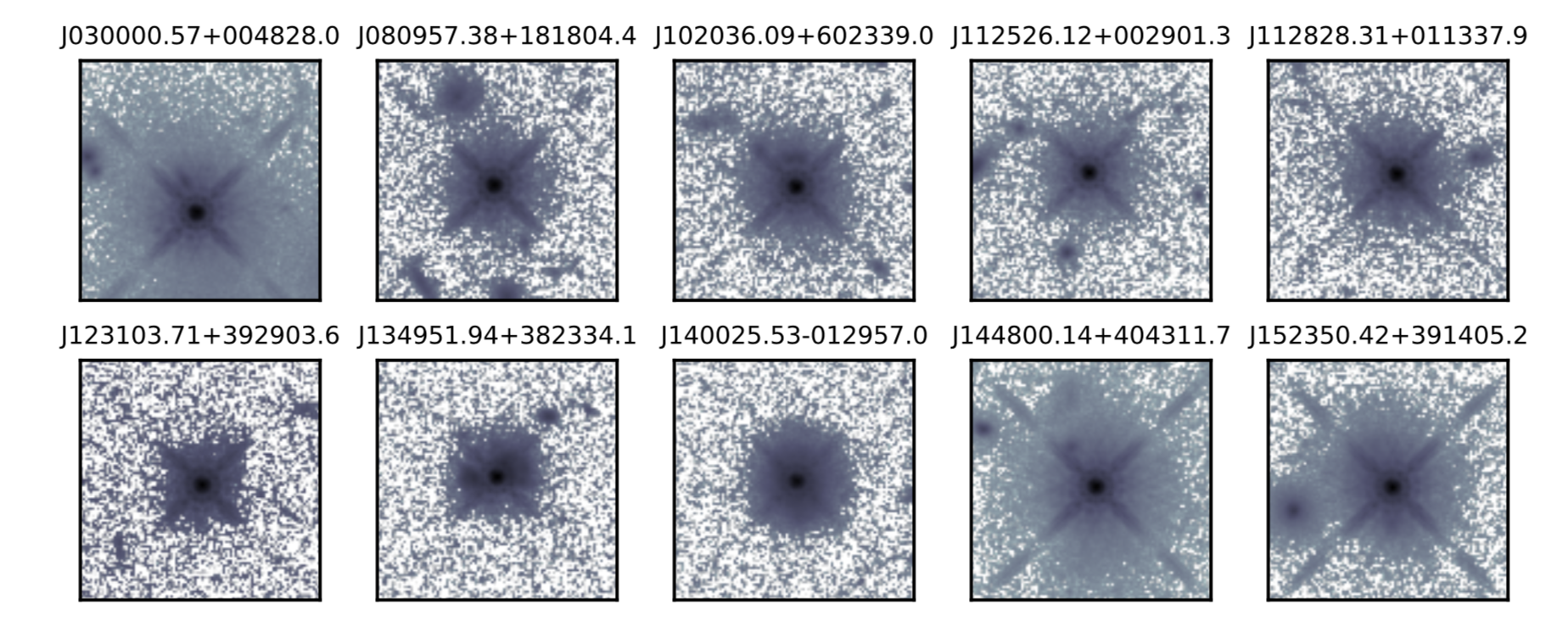}
\caption{HST F160W image cutouts of the FeLoBAL quasars (before fitting). The individual cutouts are 10\arcsec x 10\arcsec (corresponding to $\sim$ 70 kpc x 70 kpc). Note that all images are shown aligned with the detector, rather than by orientation on the sky.}
\label{fig:stamps}
\end{figure*}

\section{Sample}
\label{S:sample}

The sample includes 10 FeLoBAL quasars which were selected from a catalog of BAL quasars in the Sloan Digital Sky Survey (SDSS) by \citet{gibson_catalog_2009}, as well as the \citet{shen_catalog_2011} SDSS quasar catalog and our own additional searches through SDSS DR7 and DR10 spectra. The properties of the FeLoBAL quasars are listed in Table \ref{tab:results}.

We selected FeLoBAL quasars at the lowest possible redshifts for which FeLoBAL features are covered by SDSS spectra (0.6$<$z$<$1.1). We selected quasars with the strongest, broadest BAL absorption features, indicative of extreme powerful outflows. We also required the FeLoBAL quasars to be radio-quiet (based on non-detections in the FIRST survey). Figure \ref{fig:felobalspec} shows example spectra of two representative FeLoBAL quasars in our sample.

An important aspect of our study is comparison to reference samples of normal blue non-BAL quasars (which are older according to the evolution models). The comparison sample of normal blue quasars is taken from \citet{villforth_host_2017}. This sample consists of 20 luminous quasars (L$_{bol}\sim 10^{45-47}$ erg/s) at a redshift slightly lower than that of the FeLoBAL sample (0.5$<$z$<$0.7). \citep{villforth_host_2017} resolved 15/20 sources with absolute magnitudes -25.26 to -23.20.

While the redshift ranges for the two samples show a significant overlap, the FeLoBALs are at slightly higher redshift. A closer match could not be achieved due to the rarity of strong FeLoBALs. Differences in surface brightness dimming are minimal between the two samples, similarly, we do not expect noticable changes in merger rates across the narrow redshift range, either from theoretical consideration \citep[e.g.][]{fakhouri_merger_2010} or observations \citep[e.g.][]{bundy_slow_2004}. The difference in redshift between the sample therefore does not affect the analysis.

\section{Data}
\label{S:data}

We obtained HST WFC3/IR F160W (broad H-band) images of the 10 FeLoBAL quasars (GO Proposal 13842). Each object was observed for one orbit.

Saturation of the quasar point source can cause persistence in the WFC3/IR array in subsequent exposures. To avoid this, we used large dither steps of several arc- seconds to move any persistent quasar image away from the host galaxy in the subsequent exposure. Additionally, we used the WFC3/IR channel in MULTIACCUM mode, which takes multiple non-destructive reads during each exposure. The exposure time is determined by the sampling sequence and number of samples. We used the STEP50 sampling sequence, which initially takes several short linear reads followed by logarithmically spaced reads. The STEP50 sample sequence with 10 samples results in an exposure time of 249.3 seconds. We dithered the exposures to improve the sampling of the quasar PSF and remove detector anomalies using the WFC3-IR-DITHER-BLOB pattern with a WFC3-IR-DITHER-LINE-3PT subpattern, resulting in 9 sub-exposures and a total exposure time of 2243.7 seconds per quasar.

The individual dithered exposures for each quasar were combined using \textsc{drizzlepac} \citep{fruchter_betadrizzle:_2010,gonzaga_drizzlepac_2012}. Drizzle combination is optimized using the final scale and final pixfrac parameters. The final scale parameter specifies the size of the output pixels in arcseconds. We chose a final scale value of 0.0642\arcsec , which is half of the native WFC3/IR pixel size. The original pixels are mapped on to the output pixel grid, accounting for shifts and rotations between the individual dithered exposures. The original pixels are shrunk before being averaged on to the output pixel grid, in order to avoid convolving the image with the native pixel grid. The final pixfrac parameter specifies the amount by which the original pixel is shrunk, ranging from 0 to 1. The choice in final pixfrac is a balance between being small enough to avoid degrading the image and large enough to have uniform coverage in the combined image. We used a final pixfrac of 0.8, following \citet{villforth_host_2017}. Figure \ref{fig:stamps} shows 10\arcsec x 10\arcsec stamps for all 10 FeLoBAL quasars.

\section{Image Decomposition and Host galaxy Detection}
\label{S:psf}

\begin{table*}
\centering
\caption{Summary of FeLoBAL quasar information, \textsc{galfit} results, visual classification results, and asymmetry measurements. The columns are Name: SDSS ID; z: redshift from \citet{wild_peering_2010}; m$_{i}$: SDSS i-band magnitude; m$_{QSO}$: quasar magnitude; m$_{Galaxy}$: host galaxy magnitude; r$_{Galaxy}$: effective radius in arcseconds; Sersic: Sersic index (see Section \ref{S:psf} for a discussion on the reliability of Sersis indices); M$_{QSO}$: absolute quasar magnitude; M$_{Galaxy}$: absolute host galaxy magnitude; Class.: visual classification, see Section \ref{S:visclass} for full description, X are PSF dominated, N undisturbed, D somewhat disturbed and M clear mergers; A: asymmetry measurement. Quasar and host galaxy magnitudes are in F160W. Quantities in brackets were fixed during fitting in order to obtain a fit. J134951.94+382334.1 is best fit by two host galaxy components, of which one is offset. Both components are listed separately.}
\label{tab:results}
\begin{tabular}{lcccccccccc} 
\hline
Name & z & m$_{i}$ & m$_{QSO}$ & m$_{Galaxy}$ &  r$_{Galaxy}$ & Sersic &  M$_{QSO}$ & M$_{Galaxy}$ & Class.: & A\\
\hline
J030000.57+004828.0 & 0.900 & 16.60 & 16.47$\pm$0.04 & 18.69$\pm$0.04 &  0.93$\pm$0.01 & [1.00] &  -27.35 & -25.13 & X & --\\
J080957.38+181804.4 & 0.972 & 17.44 & 17.71$\pm$0.03 & 20.30$\pm$0.03 &  0.35$\pm$0.01 & [1.00] &  -26.32 & -23.64 & X & --\\
J102036.09+602339.0 & 0.994 & 18.25 & 17.92$\pm$0.02 & 20.21$\pm$0.02 &  0.96$\pm$0.01 & [1.00] &  -26.17 & -23.88 & D & 0.56\\
J112526.12+002901.3 & 0.864 & 17.89 & 18.18$\pm$0.02 & 20.77$\pm$0.02 &  0.62$\pm$0.02 & [1.00] &  -25.53 & -22.94 & X & --\\
J112828.31+011337.9 & 0.893 & 18.36 & 18.01$\pm$0.02 & 19.83$\pm$0.02 &  0.46$\pm$0.02 & 1.99$\pm$0.08 &  -25.79 & -23.97 & D & --\\
J123103.71+392903.6 & 1.004 & 18.52 & 18.53$\pm$0.08 & 21.13$\pm$0.08 & 0.15$\pm$0.01 &3.09$\pm$0.52  &  -25.59 & -23.97 & X & --\\
J134951.94+382334.1 & 1.094 & 18.74 & 18.77$\pm$0.01 & 20.36$\pm$0.01 & 0.53$\pm$0.01 & 0.67$\pm$0.02 &  -25.58 & -22.99 & M & 0.47\\
 &  &  &  & 21.44$\pm$0.01 &  0.34$\pm$0.01 & 0.77$\pm$0.06 &   & -22.92 &  & \\
J140025.53-012957.0 & 0.584 & 18.14 & 18.74$\pm$0.01 & 19.30$\pm$0.01 &  0.82$\pm$0.01 & 0.69$\pm$0.01 &  -23.92 & -23.66 & N & 0.18\\
J144800.14+404311.7 & 0.801 & 16.78 & 16.38$\pm$0.01 & 17.91$\pm$0.01 & 0.19$\pm$0.01 & [4.00] &  -27.13 & -25.60 & X & --\\
J152350.42+391405.2 & 0.658 & 16.42 & 16.32$\pm$0.01 & 18.91$\pm$0.01 &  0.63$\pm$0.01 & 1.41$\pm$0.04 &  -26.66 & -24.07 & D & --\\
\hline
\end{tabular}
\end{table*}

\begin{figure*}
\includegraphics[width=\textwidth]{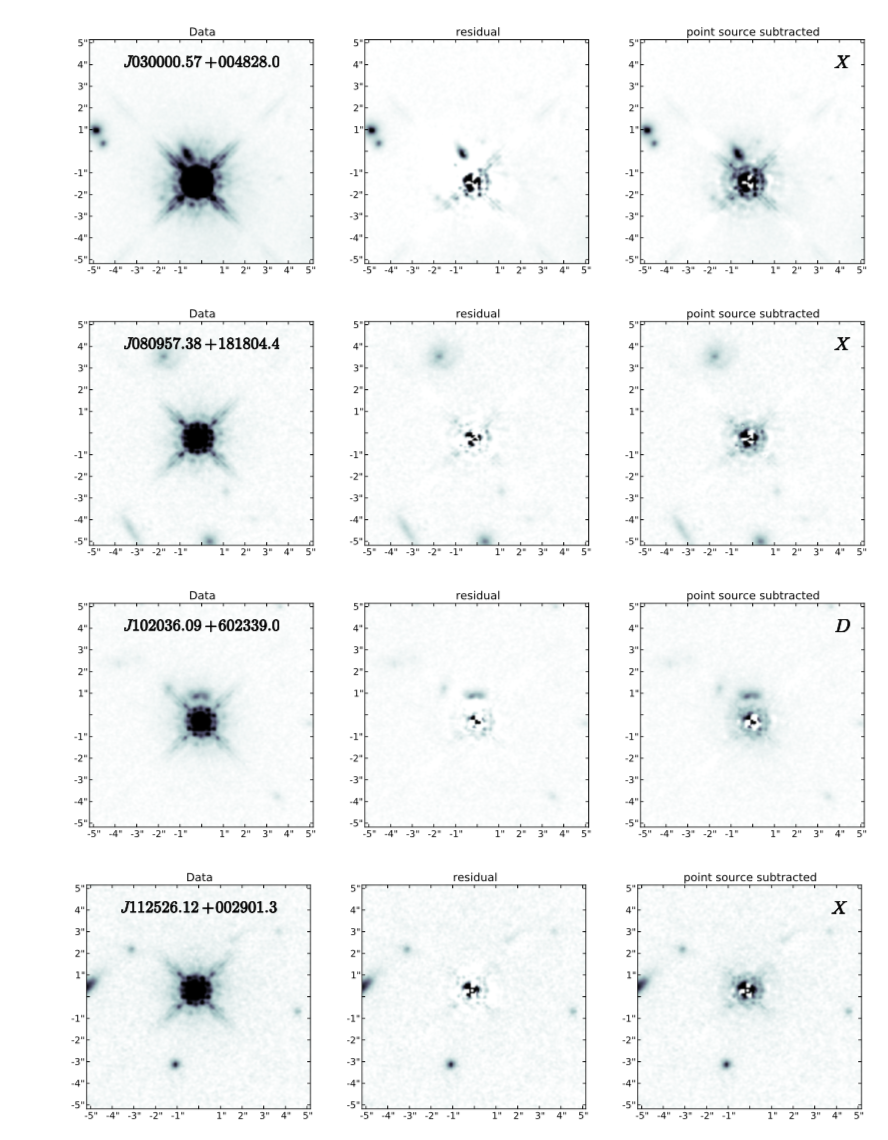}
\caption{\textsc{galfit} decomposition of the FeLoBAL quasars. Remaining sources are shown in Figures \ref{fig:fit_2} and \ref{fig:fit_3}. Image cutouts are 10\arcsec x 10\arcsec (corresponding to $\sim$ 70 kpc x 70 kpc). HST F160W images before fitting (left panel), \textsc{galfit} residuals with PSF and Sersic subtracted (middle panel), and \textsc{galfit} residuals with PSF subtracted (right panel). Visual morphological classification is indicated in the upper right corner of the right panel. Note that all images are shown aligned with the detector, rather than by orientation on the sky.}
\label{fig:fit_1}
\end{figure*}

\begin{figure*}
\includegraphics[width=\textwidth]{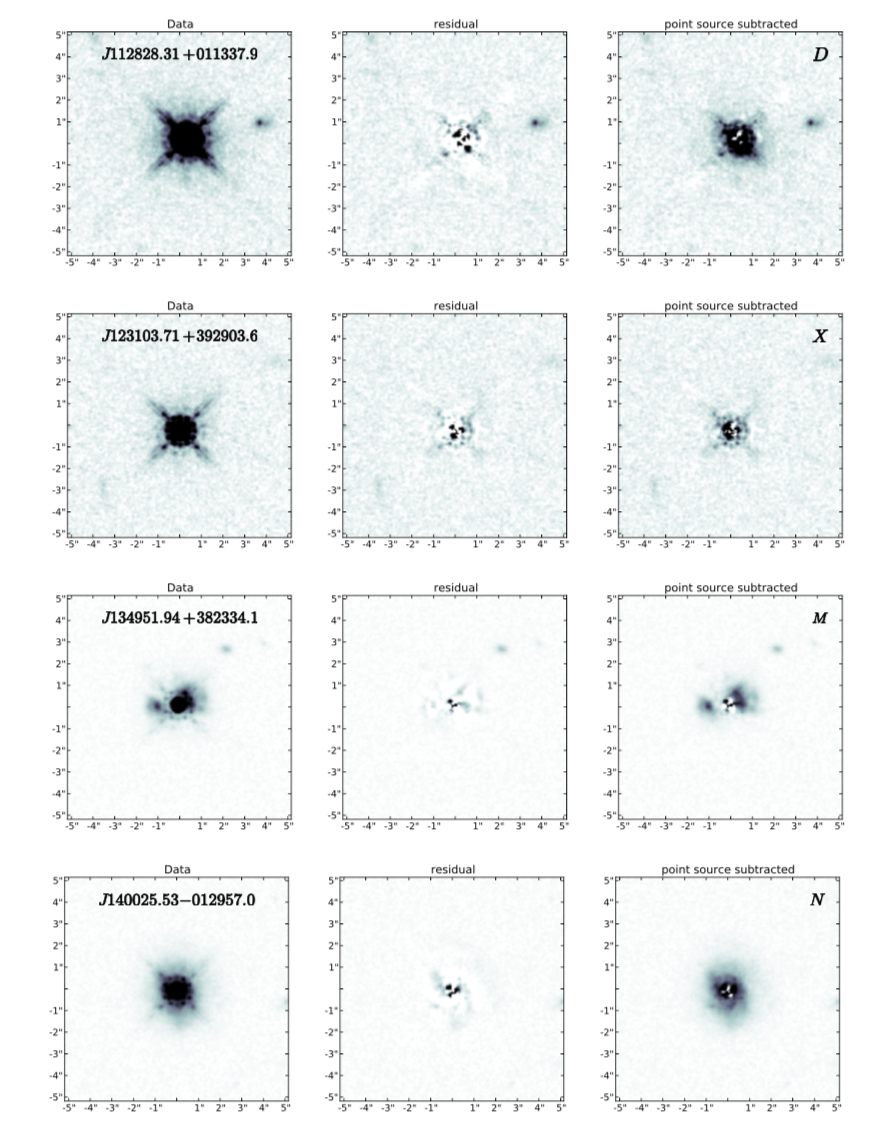}
\caption{Figure \ref{fig:fit_1} continued.}
\label{fig:fit_2}
\end{figure*}

\begin{figure*}
\includegraphics[width=\textwidth]{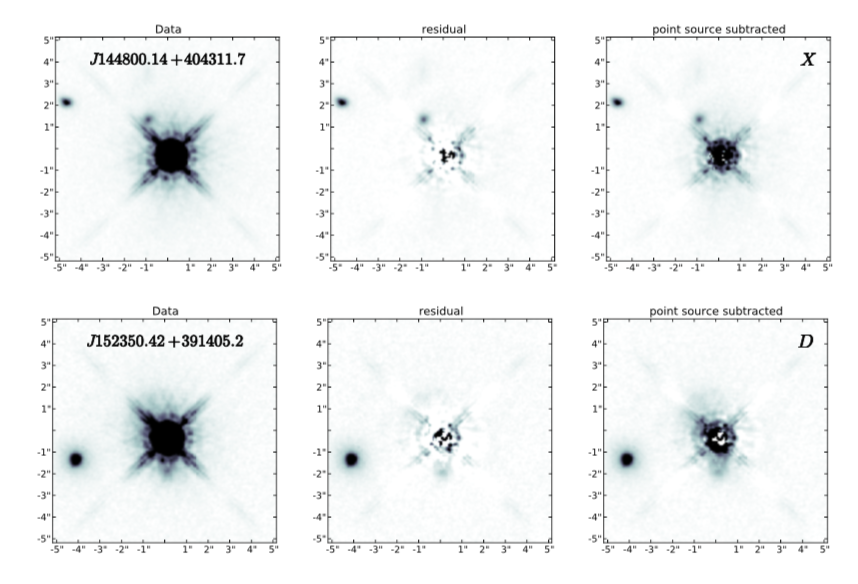}
\caption{Figure \ref{fig:fit_2} continued.}
\label{fig:fit_3}
\end{figure*}


An important part of the analysis is subtraction of the quasar nucleus to reveal the host galaxy.  Artificial point spread functions (PSFs) were created for each FeLoBAL quasar using the  \textsc{starfit} software \footnote{https://www.ssucet.org/~thamilton/research/starfit.html}, which fits PSF models to the nucleus.  The model PSF is constructed from a set of artificial point sources created with the TinyTim software \citep{krist_simulation_1995}, specifying the telescope instrument, channel, filter, source spectrum, and position on the chip.  The TinyTim PSFs are sampled on a grid nine times finer than the final pixel size, to allow for proper sub-pixel centering.  This is important, because the shape of the PSF core will look different if it is centered in the middle of a pixel, the edge, or a corner.  Telescope focus is fitted to stars in the field, avoiding errors caused by the presence of the quasar host.  \textsc{starfit} performs a non-linear, least-squares fit to the core of the quasar nucleus in the nine individual exposures (FLT files) and creates full-sized models that are then drizzled together with the same procedure used for creating the science image (DRZ file).  The final PSF model is therefore a match for the exact location and observing conditions of each object.

We used the 2D fitting algorithm \textsc{galfit} \citep{peng_detailed_2002} to simultaneously fit the quasar PSF and host galaxy. The host galaxy component is fit using a Sersic function with ellipticity and radius as free parameters, with the option of adding an additional Sersic component if needed. The \textsc{galfit} results provide the magnitude of the PSF and the magnitude, effective radius, Sersic index, axis ratio (b/a), and position angle for the Sersic component. We consider the host galaxy to be resolved when the \textsc{galfit} results do not diverge and result in physical values and the $\chi^2$ values of the fits are improved compared to PSF only fits. In cases in which the fits diverge, Sersic indices are fixed to either 1 (disk) or 4 (bulge) and the fit with the best $\chi^2$ is given. All results are presented in Table \ref{tab:results}.
In one case (J134951.94+382334.1), the host galaxy is in a clear merger, with 2 separately resolved host galaxies, in this case, we give lists the host galaxy fit results for both of the galaxies.

The FeLoBAL quasar absolute magnitudes range from -27.4 to -23.9 with a median value of -26.0. The FeLoBAL host galaxies are luminous, with absolute magnitudes ranging from -25.6 to -22.9 with a median value of -23.9). This is similar to the blue quasar host galaxies, which have absolute magnitudes $\sim$-23.5 implying stellar masses $M_* \sim 10^{10-11} M_{\odot}$ \citep{villforth_morphologies_2014,villforth_host_2017}.

The effective radii of the FeLoBAL host galaxy fits range from 0.15 to 0.96\arcsec, with a median value of 0.53\arcsec (Table 1), corresponding to $\sim$4.2 kpc at the median redshift of the sample. The Sersic indices of the host galaxy fits range from 0.67 to 3.09 (and the Sersic index was fixed at 4.00 during fitting for one of the FeLoBAL quasars), with most of the host galaxies having low (disk-like) Sersic indices.

Sersic indices can give important clues on the recent history of a galaxy, the prevalence of disk-like morphologies in our samples speaks against recent major mergers. However, the host galaxies in our sample are compact and considerably fainter than the central point source. We therefore test how reliably sersic indices can be recovered in our sample. To do so, we create simulated host galaxy quasar combinations for both disks (Sersic index 1) and bulges (Sersic index 4) for a range of magnitude differences matching the range seen in our sample. We add noise to the simulated images and then refit them using a second, slightly different, PSF to account for uncertainties in PSF constructions. We find that the magnitude of the host galaxy is recovered well down to the $\sim$3 mag differences seen in our sample. However, there are larger uncertainties in the recovered Sersic indices. While Sersic indices are recovered well for moderate magnitude differences (0.5-1 mag), fits for larger magnitude differences strongly favour low (disk-like) Sersic indices. The morphological results regarding Sersic indices are therefore likely affected by systematics.

Figures \ref{fig:fit_1}-\ref{fig:fit_3} show the \textsc{galfit} results for all 10 FeLoBAL quasars. Some merger signatures are apparent in the PSF subtracted images (right panels in Figures \ref{fig:fit_1}-\ref{fig:fit_3}), for example J134951.94+382334.1 is in a clear merger with two separately resolved hosts. Asymmetric features indicating disturbance can also be seen (e.g. J102036.09+602339.0 and J152350.42+391405.2). However these features are not common throughout the sample.

Additionally, we use the fitted PSF magnitudes to estimate the bolometric luminosities of both the FeLoBALs and blue controls. Due to the heavy absorption and obscuration, estimating the bolometric luminosities for FeLoBALs is not possible using the optical-UV magnitudes. We therefore use the absolute magnitudes in F160W/H to compare FeLoBALs and blue quasars. FeLoBAL bolometric luminosities range from $45.6 < log(L_{bol} [erg/s]) < 47.0$ with a mean of $log(L_{bol} [erg/s]) = 46.6$. The bolometric luminosities\footnote{Bolometric corrections are calculated from the SED used in \citet{hamann_extreme-velocity_2013}.} for the blue quasars range from $45.1 < log(L_{bol} [erg/s]) < 46.7$ with a mean of $log(L_{bol} [erg/s]) = 46.1$. The bolometric luminosities from the absolute F160W/H band magnitudes estimates are broadly consistent with those estimated from the X-ray and optical. The FeLoBALs therefore span the same range in bolometric luminosity, although the average of the FeLoBAL luminosities is about an order of magnitude higher. However, it should be noted that this calculations assume overall similar SEDs for the FeLoBAL and non-BAL quasars. There might be strong differences in the unmeasured FUV SEDs \citep[see e.g.][]{hamann_does_2018} that might result in large systematic uncertainties in the bolometric corrections.

Due to the spectral properties of the FeLoBALs, black hole masses are not available. It is therefore unclear if the bolometric luminosities are higher due to on average higher black hole masses or higher Eddington rates. The fact that the host galaxy magnitudes of the two samples are comparable speaks against strong differences in the black hole masses, assuming both samples follow the same M-$\sigma$ relation. The comparison of the bolometric luminosities and host galaxies luminosities therefore is most likely due to a difference in the Eddington ratios, with FeLoBALs having on average higher Eddington ratios. However, as discussed above, bolometric corrections carry large uncertainties making a detailed comparison difficult.

\section{Morphological Analysis}
\label{S:visclass}

\begin{figure*}
\includegraphics[width=\columnwidth]{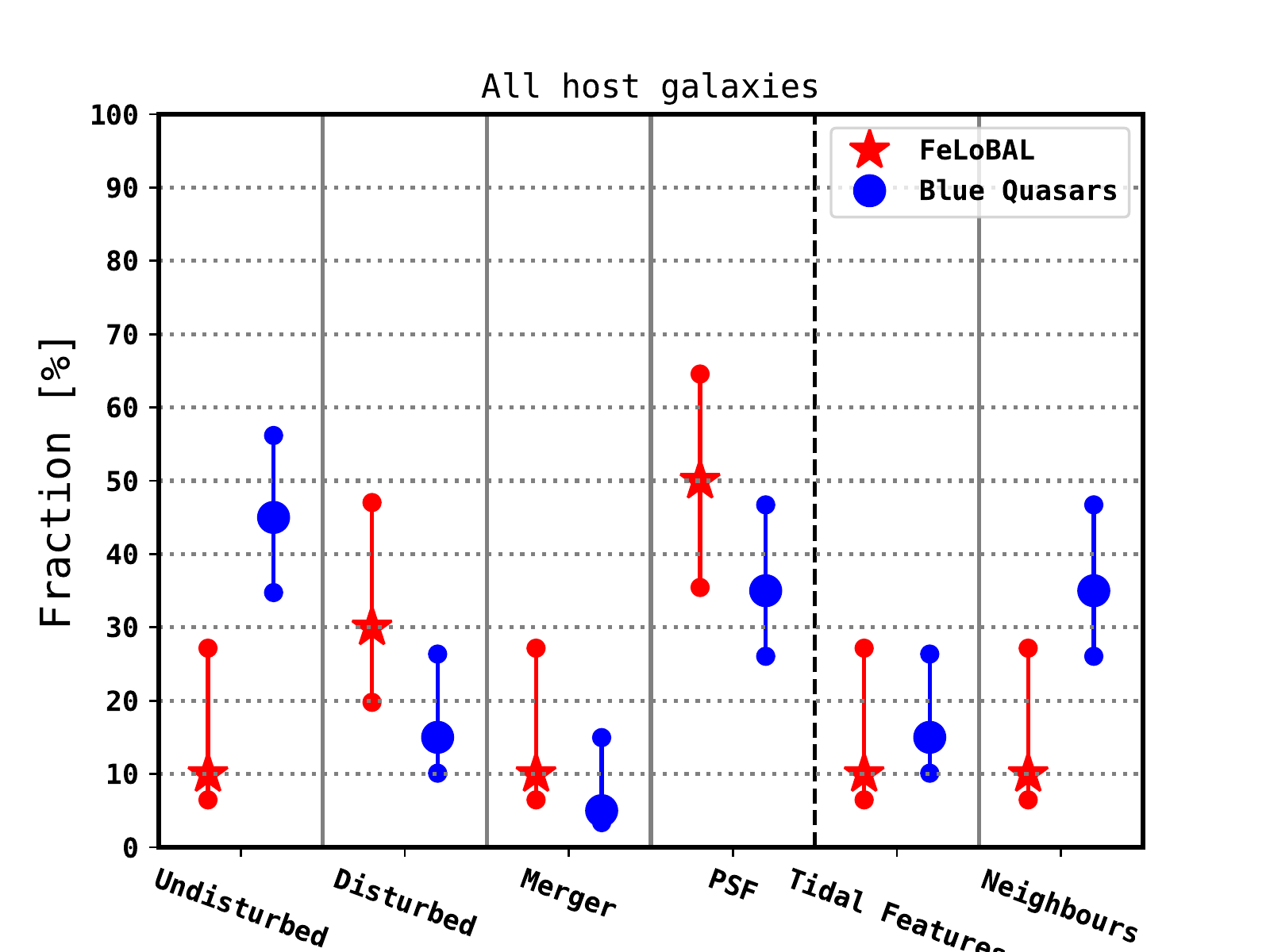}
\includegraphics[width=\columnwidth]{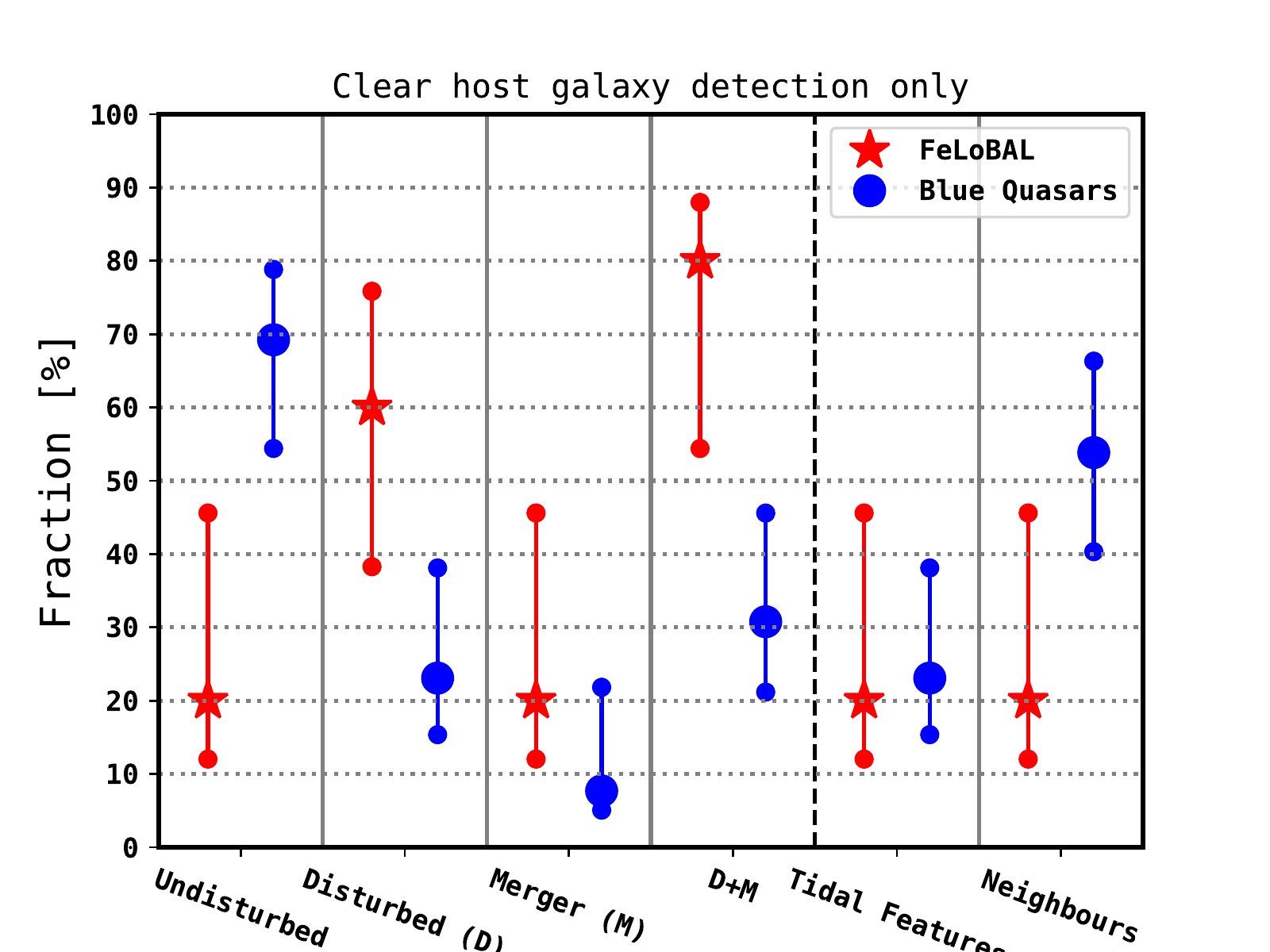}
\caption{Results of the visual classification of FeLoBAL hosts (red stars) and blue quasars (blue circles) from \citet{villforth_host_2017}. Left panel: Full sample. Right panel: Visual classification results discarding the PSF dominated sources. Error bars show 1$\sigma$ confidence intervals calculated following \citet{cameron_estimation_2011}.}
\label{fig:visclas}
\end{figure*}

We use the PSF-subtracted images of the FeLoBAL hosts as well as the blue quasar comparison sample from \citet{villforth_host_2017} for morphological analysis. We included the PSF-subtracted images from \citet{villforth_host_2017} of the 20 blue quasars in the visual classification in this study in order to compare results from the same classifiers. Note that 5 of the blue quasar hosts are unresolved, all are classified as point-source dominated. We choose to use visual classification since PSF residuals are likely to distort results from quantitative analysis.

Three human classifiers visually inspected all PSF subtracted images. The classifiers were not involved in the host galaxy fitting and we randomized the images provided to the classifiers to avoid the classifiers being able to distinguish between the FeLoBAL and blue quasars. The visual classification followed the following scheme, which was also used in \citep{villforth_host_2017}. This includes the following categories:

\begin{itemize}
\item N (Undisturbed): undisturbed
\item D (Disturbed): some signs of disturbance (asymmetric features)
\item M (Merger): clear merger (strong tidal features, double nuclei)
\item X (PSF Dominated): strong residual from point source subtraction, no classification possible
\end{itemize}

In addition to this main classification, the classifiers flagged which sources have tidal features and which sources have nearby neighbours within 10\arcsec, whether or not they appear to be interacting with the primary galaxy. To combine the classifications of the three human classifiers, we selected the majority classification. For example, when the classifications are N,N,D, the resulting classification is N. When there is no majority, i.e. the classification is N,D,M, we selected the middle ground classification of D. When two or three classifiers classified the galaxy as X, we set the classification to X. In six cases one classifier classified the galaxy as X, while the others provided classifications. In three of six cases, the other two classifiers agreed on the classification and that classification was selected. In two of six cases, the remaining two classifications were D,M and we selected D to report that two classifiers agreed there is some level of disturbance. In one case, the remaining two classifications were N,D and we selected N.

Figure \ref{fig:visclas} shows the results of the visual classification, 50$\pm15$\% FeLoBAL quasars were visually classified as PSF Dominated compared to 35$^{+12}_{-9}$\%  of blue quasars (13$^{+14}_{-4}$\% for the resolved hosts only). One FeLoBAL quasar was classified as Undisturbed (10$^{+17}_{-4}$\%), three are classified as Disturbed (30$^{+17}_{-10}$\%),  only one is classified as Merger  (10$^{+17}_{-4}$\%). All are consistent with the sample of blue quasar sample, which is consistent with the inactive galaxy population. The visual classification results are also included in Table \ref{tab:results} and presented in Figure \ref{fig:fit_1}-\ref{fig:fit_3} for each individual sources. The rates of both mild disturbances and clear mergers are consistent those seen in blue quasars.

Because of the large number of sources classified as PSF dominated, in the right panel of Figure \ref{fig:visclas} we show the results of the visual classification excluding all sources that were visually classified as PSF Dominated, noting that the PSF dominated sources are likely compact host galaxies without extended merger features (and would likely be classified as Undisturbed in the absence of the strong PSF residuals). The PSF Dominated sources are resolved with \textsc{galfit}, with relatively small effective radii (see Table \ref{tab:results} and Section \ref{S:psf}). The compactness of the FeLoBAL quasar host galaxies is also supported by the higher fraction of PSF Dominated classified sources in the FeLoBAL quasars compared to the blue quasars (left panel of Figure \ref{fig:visclas}).

After rejecting PSF dominated sources, the rates of both clear mergers and mild disturbances are higher by a factor of two in the FeLoBAL quasars compared to the blue quasar sample. For the clear mergers, the fraction of FeLoBALs classified as such is 20$^{+25}_{-8}$\% compared to 8$^{+14}_{-3}$\%. Similarly, the fraction of FeLoBALs classified as disturbed is 60$^{+16}_{-22}$\%, compared to 23$^{+15}_{-8}$\%.  No difference is observed in the fraction of objects showing clear tidal features.

Combining the disturbed and merger classes, 4/5 FeLoBALs are disturbed in some way (80$^{+8}_{-26}$\%), compared to only 4/13 in the blue quasar sample (31$^{+16}_{-9}$\%). The enhanced fraction of mergers and disturbed sources is clearly enhanced, however, due to the small number of sources, the enhancement is at low significance (4\%). The significance was calculated by drawing from the statistical distributions for the merger rates calculated following \citet{cameron_estimation_2011} and calculating the rate for which $f_{merger, FeLoBAL} > f_{merger, Blue}$. Given the currently available sample sizes, there is therefore no strong evidence for an enhancement compared to the blue quasar sample. Since this enhancement is calculated after rejecting sources due to a dominant PSF, we will discuss the significance of this result further on using simulated data.

Additionally, we quantify disturbances in the host and control galaxy morphologies using the asymmetry A measurement \citep{schade_canada-france_1995,abraham_galaxy_1996,conselice_asymmetry_2000} for those objects with weaker PSF residuals. We use these results for comparison with  \citet{villforth_host_2017}). The asymmetry is defined as:

\begin{equation}
A = \dfrac{\sum | I_{0} - I_{180}|}{2 \sum |I_0|}
\end{equation}

where $I_{0}$ and $I_{180}$ are the fluxes in the original image and the image rotated by 180$^{\circ}$, respectively. The image is rotated 180$^{\circ}$ about a center point which is chosen to minimize A. The central region of the image is masked (with a circular mask of radius up to 7 pixels based on visual inspection) to exclude the strong PSF residuals in the central pixels.

The asymmetry A measurements were hampered by the strong PSF residuals in many of the FeLoBAL quasars. In cases with strong PSF residuals (e.g. J030000.57+004828.0, right panel of Figure 3), the asymmetry measure due to the PSF residuals is clearly artificially enhanced.

Reasonable asymmetry measurements were achieved for J102036.09+602339.0, J134951.94+382334.1, and J140025.53-012957.0 (see Table \ref{tab:results}). These sources have noticeably less PSF residuals compared to the other FeLoBAL quasars (right panels in Figures \ref{fig:fit_1}-\ref{fig:fit_3}). In agreement with this, these 3 FeLoBAL quasars were not classified as PSF Dominated during the visual classification. These 3 FeLoBAL quasars have asymmetry A = 0.56, 0.47, and 0.18 (Table \ref{tab:results}). The blue quasars have asymmetry measurements ranging from 0.1 to 0.5 with median A=0.2 \citep[see Figure 6 in][]{villforth_host_2017}. The FeLoBAL asymmetries are in the range of the blue quasar asymmetries. Two of the FeLoBAL quasars have asymmetries at the high end of the blue quasar asymmetry distribution, however we are unable to do two sample tests to access the difference in asymmetry between the FeLoBAL quasars and blue quasars.

\section{Discussion}
\label{S:discussion}

Using image decomposition, we resolved the host galaxies of 10 FeLoBAL quasars. The host galaxies are luminous ($M_{gal} \sim -23.5$ in observed F160W/H) and have effective radii $\sim4.2$kpc. The fits favour disk-like morphologies, although given the large magnitude differences, we find that the Sersic indices might not be recovered reliably. In all these basic properties, they are comparable to the hosts of our comparison sample of blue quasars at slightly lower redshift from \citet{villforth_host_2017}. The rates of both clear mergers and disturbances are consistent with those seen in the blue quasars. This also means that we see no enhancement when compared to control samples, see \citet{villforth_host_2017}. We see an enhancement below statistical significance when taking only objects without strong PSF residuals into account, however, even for this subsample, only 20$^{+25}_{-8}$\% of sources show signatures of clear mergers or tidal features. The majority of disturbed sources show weaker signatures of disturbance. 

The lack of clear merger signatures in the sample suggest that the majority of FeLoBALs are not hosted by ongoing major mergers. A prevalence of disk-like morphologies is seen in our sample. While the sersic indices might not be reliable for sources with large magnitude differences, is also seen in sources with small magnitude differences. If the prevalence of disks indeed holds throughout the sample, this would also speak against recent major mergers. The tentative finding of an enhancement in disturbances weaker than a major merger could then not be explained as a sign of major mergers since the disk-like morphologies are not consistent with remnants of recent major mergers. The tentative enhancement could be due the either minor mergers that did not destroy the disk or other disturbances to the galaxy. However, as discussed above, the results on Sersic indices might not be reliable due to systematics in the fits. The hosts are therefore not inconsistent with late bulge-dominated merger remnants. 

These makes FeLoBAL hosts in this study consistent with a wide range of the literature finding quasars in mostly disk-like galaxies \footnote{Though see Section \ref{S:psf} for a discussion on reliability of Sersic indices.} showing no clear signs of recent major mergers \citep[e.g.][]{cisternas_bulk_2011,kocevski_candels:_2012,schawinski_heavily_2012,villforth_morphologies_2014,villforth_host_2017,boehm_agn_2012,gabor_active_2009,hewlett_redshift_2017,grogin_agn_2005}
This suggest that unlike extremely red quasars \citep{urrutia_evidence_2008,glikman_major_2015} or heavily obscured AGN \citep{kocevski_are_2015}, FeLoBALs are consistent with the population of host galaxies of quasars of similar luminosities. This also matches findings by \citep{violino_scuba-2_2016} who compared the star forming properties of FeLoBAL quasars to the general quasar population and found them to be consistent.

We will now compare our results to quasar populations that, like FeLoBALs, are believed to be in a transition phase of their evolution, particularly, another study of FeLoBAL hosts \citep{lawther_hubble_2018} as well as red quasars \citep{urrutia_evidence_2008}.

\citet{lawther_hubble_2018} in a recent study presented HST imaging of 4 FeLoBALs in the observed optical and IR. Their FeLoBALs span a much wider range of redshift (z$\sim$1-2). They find the hosts to be elliptical, although they note that due to the bright point sources, host galaxy fits are challenging. The colours are consistent with either dusty star formation  or old stellar populations. They find no strong disturbances, indicative of ongoing or recent major mergers, but instead find a high incidence (75$^{+10}_{-28}$\%) of nearby neighbours. This rate is higher than that seen in both our FeLoBAL and blue quasar sample (see Fig. \ref{S:visclass}), although the small number of sources means the difference is not statistically significant. Differences in redshift also make the comparison difficult. Only one source (J0300+0048) is included in both our and their sample. A neighbouring galaxy is noted in both. Due to the difference in redshift, a direct comparison to our sample is not possible. 

Another important comparison is to be made between FeLoBALs and red or heavily obscured quasars. FeLoBAL quasars show high levels of reddenning \citep{dunn_determining_2015}. Detection of FeLoBAL outflows are also common in heavily reddened sources \citep{glikman_first-2mass_2012}. This suggest a connection between obscuration and the incidence of low ionization outflows, as detected in FeLoBALs.

Red quasars and X-ray obscured quasars have been shown to have extremely high levels of clear merger signatures (\citealt{urrutia_evidence_2008,glikman_major_2015,kocevski_are_2015}, though see \citealt{schawinski_heavily_2012}). Our results for FeLoBALs
are clearly in contradiction to their results for red quasars. This suggests that while there is an overlap between the reddened and FeLoBAL population, the host galaxies of the two samples show that there are significant differences between the two populations.

\begin{figure*}
\includegraphics[width=\textwidth]{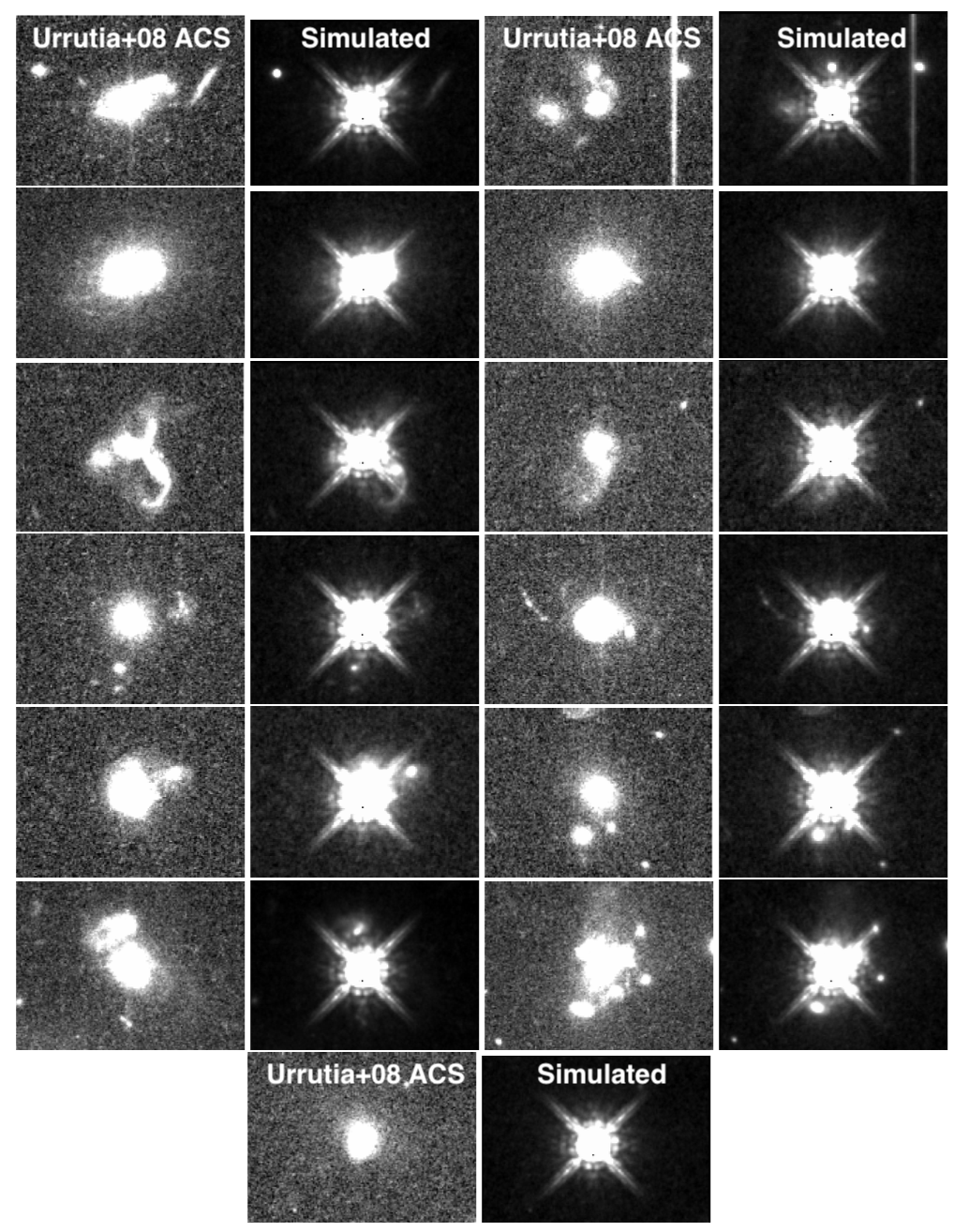}
\caption{Comparison between \citet{urrutia_evidence_2008} red quasar imaging in HST ACS F814W compared to simulated F160W matching our FeLoBAL sample}
\label{fig:redqso}
\end{figure*}

Due to differences in resolution, as well as in the contrast between central point source and galaxy, we will directly compare our results to those of \citet{urrutia_evidence_2008} and determine if differences in the observation can explain the large discrepancy. Both our FeLoBAL sample and the red quasar samples show high levels of reddening as well as strong outflows, however \citet{urrutia_evidence_2008} find 85$^{+5}_{-14}$\% of objects to show signs of strong merging, while in our sample, the fraction of clear mergers is 10${+17}_{-4}$\%. We would like to address if the differences can be explained by the different spatial resolution as well as PSF contamination. This will also be used to discuss if the full sample or the sample rejecting PSF dominated sources (see Fig. \ref{S:visclass}) gives a more realistic picture of the underlying host properties.

We alter the archival ACS F814W images from \citet{urrutia_evidence_2008} to resemble our WFC3/IR observations of the FeLoBAL quasars to explore whether the types of extended merger features seen in the \citet{urrutia_evidence_2008} red quasars would be detectable in our WFC3/IR observations if those types of features were present in the FeLoBAL quasar host galaxies. To do this, we convolve the archival ACS F814W images from \citet{urrutia_evidence_2008} with a model WFC3/IR PSF (constructed with TinyTim) to degrade the image resolution to match the resolution of our WFC3/IR observa- tions. We then add one of the FeLoBAL quasar PSF fits to the degraded images.

These altered images are shown in Figure \ref{fig:redqso}, along with the archival ACS F814W images. We do not account for the difference in galaxy flux at the wavelengths of the ACS observations ($\lambda_{rest} \sim$ 4000  \AA) versus the wavelengths of our WFC3/IR F160W observations ($\lambda_{rest} \sim$ 8500  \AA), because the galaxies are red in color (brighter at longer wavelengths), and the extended merger features are already visible in the altered images without accounting for the difference in galaxy flux at $\lambda_{rest} \sim$ 4000  \AA versus $\lambda_{rest} \sim$ 8500  \AA. Similarly, it is not necessary to fit and subtract the WFC3/IR PSF from the altered images because tidal features and asymmetric features are seen in the altered images even without PSF subtraction. For example, extended tidal tails are visible in the altered images (e.g. the third quasar in the right column of Figure \ref{fig:redqso}) and asymmetric features are also visible in the altered images (e.g. the last quasar in the right column of Figure \ref{fig:redqso}). This demonstrates that extended merger features indicative of major mergers should be detectable in the WFC3/IR observations of the FeLoBAL quasars if those type of features were present in the FeLoBAL quasar host galaxies. Differences in the observation can therefore not explain the discrepancy, the major merger rate in FeLoBALs are in fact lower than those in red quasars. These simulations also imply that the full sample gives a reasonable estimate of the merger fraction and therefore, the results using all 10 FeLoBALs rather than the smaller sample with weaker PSFs gives a realistic estimate of the host population.

\begin{figure}
\includegraphics[width=\columnwidth]{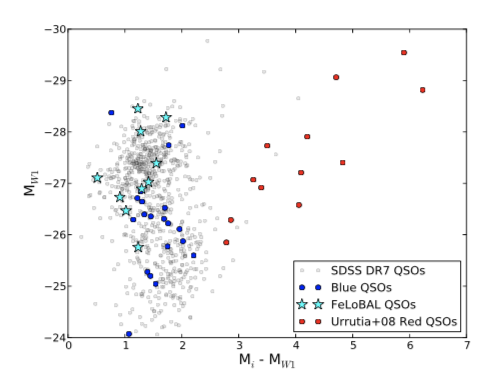}
\caption{Comparison of IR colours and absolute magnitudes between our FeLoBAL sample, the blue non-BAL sample from \citet{villforth_host_2017} and the extremely red quasars from \citet{urrutia_evidence_2008}. The figure shows the absolute magnitude in WISE 1 ($\sim$3.6$\mu$ m) observed SDSS i - Wise 1 colour.}
\label{fig:red}
\end{figure}

It should be noted however, that the FeLoBALs are less reddened than the extremely red sources in \citet{urrutia_evidence_2008}, see Fig. \ref{fig:red}. While the $m_i-m_{3.6\mu m}$ for both the FeLoBALs and blue quasars are $0.5 < m_i-m_{3.6\mu m} < 2$, those of the extremely red quasars from Urrutia range from 2-7. The FeLoBALs are redder in the optical ($m_r-m_i \sim 0.5$) compared to the blue quasars ($m_r-m_i \sim 0.2$), but are still significantly bluer than the extremely red quasars ($m_r-m_i \sim 1-2$). While FeLoBAL are reddened, they show considerably lower levels of reddening than extremely red quasars.

An interesting difference between the FeLoBAL and red quasar samples is that our sample was selected to be radio-quiet, while the extreme red quasar sample studied in \citep{urrutia_evidence_2008} is radio-loud. Radio-loud AGN are more often found to be associated with mergers \citep{ramos_almeida_are_2011,chiaberge_radio_2015}. This might suggest that there is a strong difference between radio-loud and radio-quiet quasars samples, although earlier studies  find no such differences in direct comparisons \citep{dunlop_quasars_2003}.

Some theoretical studies have suggested that major mergers should dominate the highest luminosity AGN \citep[e.g.][]{hopkins_characteristic_2009,somerville_semi-analytic_2008}. Both \citet{villforth_morphologies_2014} and \citet{villforth_host_2017} show no such trend when compared to control, while \citet{treister_major_2012} find a trend using a heterogeneous compilation of literature data. Given the simulations presented above, we showed that our full morphological sample can be used to determine merger rates, our merger rates (10$^{+17}_{-4}$\%) as well as disturbed rates (30$^{+17}_{-10}$\%) are below the updated luminosity merger rate relation by \citet{glikman_major_2015}. The expected merger rates from \citet{glikman_major_2015} at the bolometric luminosity of our sample ( $log(L_{bol} [erg/s]) = 46.6$) are $\sim70\%$. Taken the our merger rates as a comparison point, the difference is therefore technically statistically significant (p$<$0.1\%). Our sample is too small and affected by PSF residuals to study a luminosity evolution within the sample. However, we note that comparing merger rates between different samples is not straighforward given different methodologies used as well as different sensitivities to merger features in different datasets either due to image depth or wavelength used. This study does therefore not support the trend of major merger fraction with AGN luminosity seen in \citet{treister_major_2012} and \citet{glikman_major_2015}.

FeLoBAL quasars in our sample are therefore not consistent with the extremely red quasar population, instead, their host galaxies most closely match those of normal blue quasars. This suggests that while some red quasar samples have high incidences of powerful outflows, the outflows themselves are not strongly linked to major mergers. The major merger rate is also not enhanced, despite the high luminosities of the FeLoBAL quasars studied here. Our results are therefore inconsistent with recent claims of a dependence of the merger rate on quasar luminosity.

\section{Conclusions}
\label{S:end}

We analyzed WFC3/F160W (H band, $\lambda_{rest}\sim8500\AA$) imaging of 10 FeLoBAL quasars at z$\sim$0.9, performed image decomposition to analyze the host galaxies and performed morphological analysis. Our results can be summarized as follows:

\begin{itemize}
\item The host galaxies were resolved for all sources in the sample. The hosts are luminous $-25 > M_{host} > -22$, compact ($\sim$4 kpc) and predominantly disk-like, although results on Sersic indices might carry systematic errors. They are of comparable luminosity to the hosts of luminous blue quasars at similar redshift and luminosity.
\item The bolometric luminosities of the FeLoBALs   range from $45.6 < log(L_{bol} [erg/s]) < 47.0$ with a mean of $log(L_{bol} [erg/s]) = 46.6$, higher than those of blue quasars ($45.1 < log(L_{bol} [erg/s]) < 46.7$ with a mean of $log(L_{bol} [erg/s]) = 46.1$, calculated using the same method). Given the similarities in host galaxy absolute magnitudes, this difference could be due to higher Eddington ratios in the FeLoBAL sample.
\item In the morphological analysis, 30$^{+17}_{-10}$\% of FeLoBAL hosts show low levels of disturbance and an additional 10$^{+17}_{-4}$\% show clear signs of merging. 50$^{+15}_{-15}$\% could not be clearly classified due to clear PSF residuals. While the raw rates of disturbance are comparable to those of blue quasars with resolved hosts, taking only the objects without strong PSF residuals into account, the rate of both low level disturbances and clear mergers are twice as high in the FeLoBALs compared to blue quasars. When combining both mergers and disturbed sources, the merger rate is enhanced (80$^{+8}_{-26}$\% in FeLoBALs, compared to 31$^{+8}_{-26}$ for the blue quasars. The excess in the merger + disturbed rate has a significance of 4\%.
\item Quantitative morphological analysis proved challenging due to the strong PSF residuals. The asymmetry A could be calculated for three sources, all of which showed asymmetries broadly consistent with the comparison sample of blue quasars.
\item The FeLoBAL host galaxies are not compatible with the extreme mergers seen in heavily reddened quasars \citep{urrutia_evidence_2008}. Even after taking differences in resolution and point source contamination into account, FeLoBALs show considerable lower levels of mergers. This suggest that while there is a clear overlap between reddened quasars and FeLoBALs, the host galaxies show considerable differences in the hosts of the two samples.
\end{itemize}

Our results are therefore consistent with a picture in which FeLoBAL hosts at moderate redshifts ($z\sim0.9$) are only insignificantly more disturbed than blue quasars. The differences to red quasar samples  show that the association between major mergers and red quasars is not mirrored in objects with extreme outflows. We detect a tentative enhancement of milder disturbances in FeLoBALs when compared to blue quasars, but this is not statistically significant. Future studies will require both larger samples to determine if differences seen are significant, as well as comparison between samples of quasars taking into account a wide range of properties, such as luminosity, radio-loudness, obscuration in gas and dust as well as the presence of outflows.

\section*{Acknowledgements}

We thank the referee for constructive and helpful comments. This work is based on observations made with the NASA/ESA HST, obtained from the data archive at the Space Telescope Science Institute. STScI is operated by the Association of Universities for Research in Astronomy, Inc. under NASA contract NAS 5-26555. HH and FH acknowledge support from the USA National Science Foundation grant AST-1009628.

\bibliographystyle{mnras}
\bibliography{felobal} 

\bsp	
\label{lastpage}
\end{document}